\documentstyle[aps,prb]{revtex}

\begin{document}

\twocolumn[\hsize\textwidth\columnwidth\hsize\csname 
@twocolumnfalse\endcsname

\title{Field-dependent Vortex Pinning Strength in a Periodic Array of Antidots}
\author{A.V. Silhanek, S. Raedts, M. Lange, and V.V. Moshchalkov}

\address{Laboratorium voor Vaste-Stoffysica en Magnetisme,\\
K.U.Leuven, Celestijnenlaan 200D, B-3001 Leuven, Belgium.}

\date{\today}
\maketitle

\begin{abstract}
We explore the dynamic response of vortex lines in a Pb thin film with a periodic array of antidots by means of ac-susceptibility measurements. For low drive field amplitudes, within the Campbell regime, vortex motion is of intra-valley type and the penetration depth is related to the curvature of the pinning potential well, $\alpha$. For dc-fields below the first matching field $H_1$, $\alpha$ reaches its highest value associated with a Mott Insulator-like phase where vortex lines are strongly localized at the pinning sites. For $H_1<H_{dc}<H_2$, the response is mainly due to the interstitial vortices and $\alpha$ drops to smaller values as expected for this metallic-like regime. Strikingly, for $H_2<H_{dc}<H_3$, we observe that $\alpha$ reduces further down. However, for $H_3<H_{dc}<H_4$, a reentrance in the pinning strength is observed, due to a specific configuration of the flux line lattice which strongly restricts the mobility of vortices. We present a possible explanation for the measured $\alpha(H_{dc})$ dependence based on the different flux line lattice configurations.
\end{abstract}

\pacs{PACS ........}
\vskip1pc] \narrowtext

Aligned linear defects strongly influence the static and dynamic behavior of vortex lines in the mixed state. It has been extensively demonstrated that the introduction of columnar defects by heavy ions irradiation gives rise to several vortex phases and a considerable enhancement of the critical current.\cite{blatter,columnar} Theoretically, the expected vortex behavior should change significantly when the number of vortices exceeds the number of columnar defects. However, the disorder in the randomly distributed pinning sites together with the pinning energy dispersion of the tracks lead to the appearance of matching effects as washed out transitions or just barely defined crossovers.\cite{suppression} In contrast to that, the introduction of nanoengineered periodic arrays of artificial pinning centers, where both topological and energetic disorder are absent, provides an ideal model system to explore the matching transitions and study the ``zoology'' of the vortex lattice configurations.\cite{review} 

In superconductors with periodic pinning arrays (PPA) the pinning properties are considerably improved at the applied field $H_{dc}=H_1$, where the density of vortices equals the density of pins.\cite{review} This behavior is also present at higher commensurability fields $H_n = n H_1$, with $n$ an integer number. However, the enhanced pinning progressively diminishes with increasing field since the vortex-vortex interaction becomes more relevant. An additional field dependence of the critical current arises due to rearrangements of the vortex lattice configuration itself.\cite{review}

Most of these vortex configurations have been directly imaged by Lorentz and scanning Hall Probe microscopies,\cite{harada,hall} and Bitter decoration\cite{decoration}. However, none of these techniques allow one to extract the pinning potential depth. In order to do that, it is necessary to use indirect techniques like transport, dc-magnetization or ac-susceptibility. In transport and dc-magnetization measurements, the system is driven out of equilibrium, thus allowing the determination of critical currents and pinning strength of a moving flux line lattice (FLL), distinct of that observed by means of vortex imaging techniques. In contrast, low amplitude ac-susceptibility measurements permit one {\it to probe the effective pinning potential without modifying the vortex distribution}, thus making possible a direct correlation between FLL images and the observed dynamical response.

Indeed, several theoretical and experimental works\cite{krusin,pasquiniPRB} indicate that the ac response of a system with unidirectional defects exhibits, for small values of the alternating field drive $h_{ac}$, a linear behavior characterized by an ac-susceptibility $\chi$ independent of $h_{ac}$ and a very small dissipation. In this regime, known as the {\it Campbell Regime},\cite{campbell} where vortices oscillate inside the pinning centers, it is possible to estimate quantitatively the strength of the pinning potential by analyzing the field dependence of the ac-susceptibility.

In this work, we present measurements of the ac-susceptibility $\chi=\chi^{\prime} + i \chi^{\prime \prime}$ of a Pb thin film with a square periodic array of holes (antidots). We mainly concentrate on the linear regime where vortex displacement never surpasses the pinning range and the technique can be thought of as a passive non-destructive method to measure the pinning potential. From the field dependence of the ac-penetration depth, $\lambda$, we determine the curvature of the pinning potential well, $\alpha$, for different matching fields. This allows us to estimate quantitatively, for the first time, the pinning potential for each specific vortex configuration. In particular, we show that the strength of the effective pinning potential exhibits a reentrant behavior for $H_3<H_{dc}<H_4$, where vortex mobility is strongly restricted.

The studied sample was a 50-nm-thick Pb film of rectangular shape (1.8 mm $\times$ 3 mm), with a square antidot lattice of period $d=1 \mu m$, which corresponds to the first matching field $H_1=\Phi_0/d^2=$20.7 Oe, with $\Phi_0$ the superconducting flux quantum. The antidots have a square shape with a size $b =$ \mbox{0.25 $\mu$m}. The maximum number $n_s$ of flux quanta that an antidot can hold is approximately given by $n_s \approx b/4 \xi(T)$,\cite{schmidt} where $\xi(T)$ is the superconducting coherence length. For most of our samples we have estimated $\xi(0) \approx 38 nm$ and considering the usual Ginzburg-Landau temperature dependence we obtain $n_s \approx 0.6$ for $T=$6 K, the lowest temperature studied. Thus, we rule out the possibility of multiquanta occupation in a single hole.

The film was electron beam evaporated in a molecular beam epitaxy apparatus onto liquid nitrogen cooled SiO$_2$ substrates with a predefined resist-dot pattern and after the resist was removed in a lift-off procedure, the sample was covered with a protective Ge layer of 20 nm.\cite{review} The sample has a T$_c$=7.2 K with a transition width of 0.2 K, for $H_{dc}$=0 and $h_{ac}$=0.5 Oe. The ac-measurements were carried out in a commercial Quantum Design-PPMS device with drive field amplitudes $h_{ac}$ ranging from 88 mOe to 10 Oe, and the frequency $f$ from 10 Hz to 15 kHz. In all cases, the data were normalized by the same factor corresponding to a total step $\Delta \chi^{\prime} =$ 1, with $H_{dc}=$0.


In order to identify the linear regime, we performed a series of measurements of $\chi=\chi^{\prime} + i \chi^{\prime \prime}$ as a function of the ac excitation $h_{ac}$ for several temperatures at fixed $f=$3837 Hz and $H_{dc}=$20.7 Oe. In \mbox{Fig. 1} we show one of these measurements for $T=$6.4 K. The Campbell regime is characterized by a change in the magnetization $dM=\chi h_{ac}$ proportional to $h_{ac}$ together with a scarce dissipation. We observe that $\chi h_{ac}$ departs from a linear response at $h_{ac} \sim$ 1 Oe. We also note that while in the linear regime the dissipation is negligible, an increase in the ac losses appears at the onset of nonlinearity. As we will show below, the range of ac-fields where $\chi^{\prime \prime} \sim$ 0, is a reliable criteria to determine the range of the validity of the linear regime.

An alternative way for determining the Campbell regime comes from the study of higher components of $\chi$. Indeed, nonlinearities should manifest themselves as higher harmonics $\chi_m$ with $m$ being an odd number.\cite{clem-sanchez} We confirm this by measuring up to 9 components of $\chi$. We find that for the lowest $h_{ac}$, $\chi_m \approx$0 for $m>$1, while for $h_{ac}>$1 Oe, $\chi_m \neq$ 0 for $m$=3,5,7 and 9. The acquisition of these components allows us to build the magnetization curve as a function of $h_{ac}$ for two extreme excitations $h_{ac}$ at $T=$7 K and $H_{dc}=$10 Oe, as is shown in the inset of \mbox{Fig. 1}. We clearly observe that for $h_{ac}=$0.09 Oe the response is Meissner-like and almost free of dissipation ($\chi^{\prime \prime} \approx 0$). In contrast, for $h_{ac}=$3 Oe the response can be satisfactorily described using the Bean Critical State model and the losses, given by the area of the loop, are no longer small. According to the Bean model for a disk of radius $R$ and thickness $\delta$, the saturation magnetization $M_{sat}$ is related to the characteristic penetration field $H_p$, through $H_p = 3 M_{sat} \delta / 2 R$.\cite{clem-sanchez} For the data showed in the inset of \mbox{Fig. 1}, 4$\pi M_{sat} \sim$ 1400 Oe and thus 2$H_p \sim$ 1.6 Oe, in good agreement with the field needed to invert the critical state profile. 


Since the limits of the linear regime are determined mainly by the strength of the pinning sites, it is expected that the range in which this regime exists decreases with both, field and temperature, for a fixed $h_{ac}$. This behavior becomes evident in \mbox{Fig. 2}, where the out-of-phase component $\chi^{\prime \prime}$ (upper panel) and the in-phase component $\chi^{\prime}$ (lower panel) are shown as a function of the scaled field $H_{dc} / H_1$ for two values of $h_{ac}$ at $T=$7 K. 

For small amplitudes ($h_{ac}=$0.09 Oe), and for $0<H_{dc}<H_1$ a high screening and very small losses are observed. However, for $H_{dc}>H_1$, the screening exhibits an overall decrease together with a sequence of minima at $H_{dc}=H_n$ caused by local depressions in the losses. This behavior can be understood within a scenario based on the picture of vortices oscillating inside their pinning sites, without large excursions. Indeed, at the commensurability fields, where the effective pinning is maximum, a higher screening and a reduction of dissipation due to the more confined movement of vortices, is expected.

On the other hand, at $h_{ac}=$1 Oe, deep within the critical state, the response is remarkably different. Although the local dips at $H_{dc}=H_n$ in the screening curve are still present, now these minima are accompanied by a corresponding peak in the losses. This is a fingerprint of the critical state regime.\cite{metlushkoPRB} At this high amplitude value, vortices are no longer localized in small pinning valleys, but instead they can travel larger distances (inter-valley motion). As we pointed out above, in this regime, losses are given by the area of the $M h_{ac}$-hysteresis loop (see inset of \mbox{Fig. 1}) whose width is determined by the critical current of the system. At $H_{dc}=H_n$, the critical current reaches a maximum and therefore also the losses are maximized, as is indeed observed in our experiments. Based on the previous analysis we will adopt as a conservative and safe criterion to ensure a linear regime, the limit of the used $H_{dc}$ range, to the zone where $\chi^{\prime \prime} <$0.01. 

In the upper panel of \mbox{Fig. 2} we have also included $\chi^{\prime \prime}(H_{dc})$ curves for $T=$6.8 K and 6.5 K, at $h_{ac}=$0.09 Oe. As expected, we find that with decreasing $T$, the range where the system is free of dissipation, becomes broader. In particular, at $T=$6.5 K, the losses are negligible for the whole field range studied.

From now on, we will concentrate on the field dependence of $\chi$ in the linear regime for $h_{ac}=$0.09 Oe, at two different temperatures, $T=$6.5 K and 6.8 K. In this regime, the ac magnetic response is determined by the complex penetration depth $\lambda(B \sim H_{dc},T,f)=\lambda_R+i\lambda_I$. For ideal systems without dissipation, $\lambda$ is real and $\chi^{\prime \prime} = 0$. However, dissipation leads to a $\chi^{\prime \prime} \neq 0$ and a ratio $\epsilon=\lambda_I/\lambda_R$, although small, is finite. According to ref.\cite{brandt}, $\chi^{\prime}$ is solely determined by $\lambda$ and the sample geometry through $\chi \approx -1+ \sum_{n}^{20} [c_n/(\Lambda_n+\varphi)]$, where $\varphi=C/\lambda^2$, $C$ is a constant depending on the sample shape, and $\Lambda_n$ and $c_n$ are real numbers. We can invert this equation in order to compute $\lambda$ from our $\chi$ data, as described in ref.\cite{pasquiniPC}.

For our experimental conditions $\epsilon \ll 1$ and $\lambda^2 \approx \lambda_R^2(B,T) = \lambda_L^2(T) + \lambda_C^2(B,T)$, where $\lambda_L(T)$  and $\lambda_C(B,T)=\sqrt{B\Phi_0/4\pi \alpha}$ are the London and Campbell penetration depths, respectively. The Labusch constant, $\alpha$, represents the curvature of the pinning well assumed to be parabolic for small displacements.

\mbox{Fig. 3} shows $\lambda_R^2$ as a function of the reduced field $H_{dc}/H_1$ for $T=$6.5 K. The most evident feature of this figure is that for well defined field ranges (0 $\leq H_{n-1} < H_{dc} < H_n \leq H_5$), $\lambda_R^2$ follows a linear field dependence, $\lambda_R^2 \propto B/\alpha_n$, in agreement with the Campbell regime scenario, with a Labusch parameter $\alpha_n$ clearly depending on the field range. This is one of the main results of this work. The same behavior is observed at $T=$6.8 K (see upper inset of \mbox{Fig. 3}) although, as we previously noticed, the linear regime is, in this case, constrained to $H_{dc} < H_3$.

Since at $H_{dc}=$0, $\lambda_R^2 = \lambda_L^2(T)=\lambda_L^2(0)/(1-t)$, where $t=T/T_c$, using the data of \mbox{Fig. 3} we can easily estimate the London penetration depth at $T=$0. The absolute value $\lambda_L(0) \approx $52 nm so obtained, turns out to be relatively close to the accepted value $\lambda_L(0) \approx $ 47 nm.\cite{lambda}

In all cases, $\lambda <$ 1$\mu$m, and therefore the penetration of the compressional vortex waves involves just a few units cells of the antidots array from the sample border. As soon as a new vortex configuration develops right after a commensurability field $H_n$, with a consequent change in the average pinning strength, it will be simultaneously detected in the ac susceptibility response. Since our field step is 1 Oe, the density of the new developing FLL configuration is high enough to record a sharp change at $H_n$.

How actually the new vortex configuration penetrate in a sample with PPA has been a matter of controversy during the last years.\cite{cooley,reichhardt-97} Cooley and Grishin\cite{cooley} have shown that in a one-dimensional system with PPA, a terraced flux profiles should be established. In our system (thin films), due to the small gradient in the flux profile, a single terrace critical state is expected. According to this model, increasing the field beyond $H_n$, leads to the motion of the new vortex configuration towards the center of the sample. Since we are sensing the magnetic response at the sample border, the same $\lambda^2$ should be measured as $H_{dc}$ increases and therefore no field dependence should be observed. This is in sharp contrast with our observations. Instead of that, we argue that for each applied field a mixture of the patterns corresponding to the nearby matching fields distributes in the whole sample.\cite{reichhardt-97} Within this scenario, $\alpha$ so obtained results from an average of the different pinning potentials associated with each existing vortex configurations.

In the lower inset of \mbox{Fig. 3} we show a zoom of the same data as in the main panel, for $H_{dc} < H_2$. We observe that for $H_{dc} < H_1$, where vortices are strongly pinned by the antidots, $\alpha$ is larger than for $H_{dc} > H_1$ where caged interstitial vortices, much weaker pinned and with higher mobility, dominate the ac response. The Labusch parameter obtained from these slopes is summarized in \mbox{Fig. 4} for all fields studied. The sketches of different stable vortex configuration according to previous vortex imaging experiments\cite{harada} and simulations,\cite{configurations} are also shown.

Since for $H_{dc} > H_2$ the entity which respond to the external ac excitation is no longer a well identified single vortex, but a complex movement involving all interstitial vortices in a unit cell, the interpretation of the ac susceptibility for these higher fields becomes more puzzling. For instance, at $H_{dc} = H_2$, vortices start to form interstitial dimers which can adopt two different orientations separated by a small energetic barrier. When the drive field is applied, a librational motion of the vortex pairs is excited. Since the observed Labusch parameter is even smaller than for $H_{dc} < H_2$ we argue that this motion permits (in average) larger excursions of the vortex lines than the movement of single interstitial vortices at the same $h_{ac}$. 

For fields higher than $H_3$, the unit cell motif consisting of a rhombic configuration, loses the rotational degree of freedom and consequently, a reentrance of $\alpha(H_{dc})$ is observed. On top of that, because of the system is now changing from a phase with higher mobility to another phase with lower mobility, the density of the new unit cell motif needed for significantly change the measured $\alpha$, should be larger. As a consequence, the transition does not occur exactly at $H_3$ but a field slightly higher, as we indeed observe.

For fields $H_{dc} > H_4$, vortices in a unit cell distribute forming a square arrangement where the rotational degree of freedom is recovered, thus giving rise to an easy librational movement and a lower average pinning potential. At fields higher than the fifth matching condition nonlinearities develop and no reliable Labusch constants can be determined. It is important to stress that the linear dependence of $\lambda_R^2$ with field for $H_{dc} > H_2$ is a highly non trivial result. Clearly, the complex mechanisms involved during the ac excitation deserves further experimental and theoretical investigation.

In general, the main trends shown in Fig. 4 are similar to those reported in simulation studies made recently by Reichhardt et al.\cite{reichhardt} In that work, the authors determine the dependence of the critical depinning force, as a function of $H_{dc} / H_1$ for a system with a square pinning array. In agreement with our experimental observations they find a series of roughly constant plateaus of decreasing value. However, unlike the Labusch constant, the depinning force density and the critical current depend on the depth of the pinning potential (rather than its curvature) as well as the FLL configuration, and therefore this analogy should not be pushed too far. Essentially transport and ac-measurements in the linear regime are not directly comparable.

Finally, it is worth noting that in similar studies performed in high temperature superconductors with columnar defects introduced by irradiation,\cite{pasquiniPRB} the change in the slope at the first matching condition turned out to be a blur crossover with an onset around $H_1 /$2, presumably due to the topological and energetic disorder of these defects.\cite{suppression} In those studies a field dependent Labusch parameter was found for $H_{dc} > H_1 /$2, as a result of both, a disordered state where a single unit cell motif can not be identified, and the lattice interactions when the system approaches to the collective regime. In our system with PPA where vortices of same species have the same surrounding, the regimes are connected by abrupt transitions and $\alpha$ remains constant for the low fields studied.

In summary, we have used the ac-susceptibility technique as a non-disturbing tool for determining the influence of the vortex lattice arrangement in the averaged pinning strength. We have estimated quantitatively the value of these effective pinning potentials for several stable vortex configurations. This powerful method has the main advantage of an easy extension to different systems with periodic pinning array and the possibility of a direct comparison of the pinning energies found from different experiments.

We would like to thank M.J. Van Bael for helpful discussions and R. Jonckheere for fabrication of the resist pattern. This work was supported by the Belgian Interuniversity Attraction Poles (IUAP), Flemish Concerted Action Programs (GOA), the Fund for Scientific Research Flanders (FWO) and ESF ``VORTEX'' program.

\bibliographystyle{prsty}

\begin{figure}
\caption[]{{\small Drive dependence of $dM=\chi h_{ac}$ at the first matching field $H_1$, $T=$6.4 K and $f=$3837 Hz. The arrows indicate the onset of the nonlinear regime. The inset shows the normalized magnetization derived from several harmonics $\chi_n$ for two extreme amplitudes $h_{ac}=$0.09 Oe (filled symbols) and $h_{ac}=$3 Oe (open symbols), at $T=$7 K.}}
\label{fig1}
\end{figure}

\begin{figure}
\caption[]{{\small Field dependence of the ac-susceptibility $\chi=\chi^{\prime} + i \chi^{\prime \prime}$ components for $h_{ac}=$0.09 Oe (filled circles) and $h_{ac}=$1 Oe (open circles) at $T=$7 K. In the upper panel is also shown $\chi^{\prime \prime}$ for $T=$6.5 and 6.8 K, at $h_{ac}=$0.09 Oe.}}
\label{fig2}
\end{figure}

\begin{figure}
\caption[]{{\small Field dependence of $\lambda_R^2$ at $T=$6.5 K (main panel) and $T=$6.8 K (upper left inset), for $h_{ac}$=0.09 Oe. The lower right inset shows a low field zoom of the main panel data. The horizontal and vertical units in the insets are the same used in the main panel. The lines are guide to the eye.}}
\label{fig3}
\end{figure}

\begin{figure}
\caption[]{{\small Field dependence of the Labusch parameter $\alpha$ in units of its maximum value $\alpha^{max}$ at $T=$6.5 K. The sketches represent the different stable vortex configurations where empty antidots are drawn as open circles, occupied antidots as big filled circles and interstitial vortices as small filled circles.}}
\label{fig4}
\end{figure}

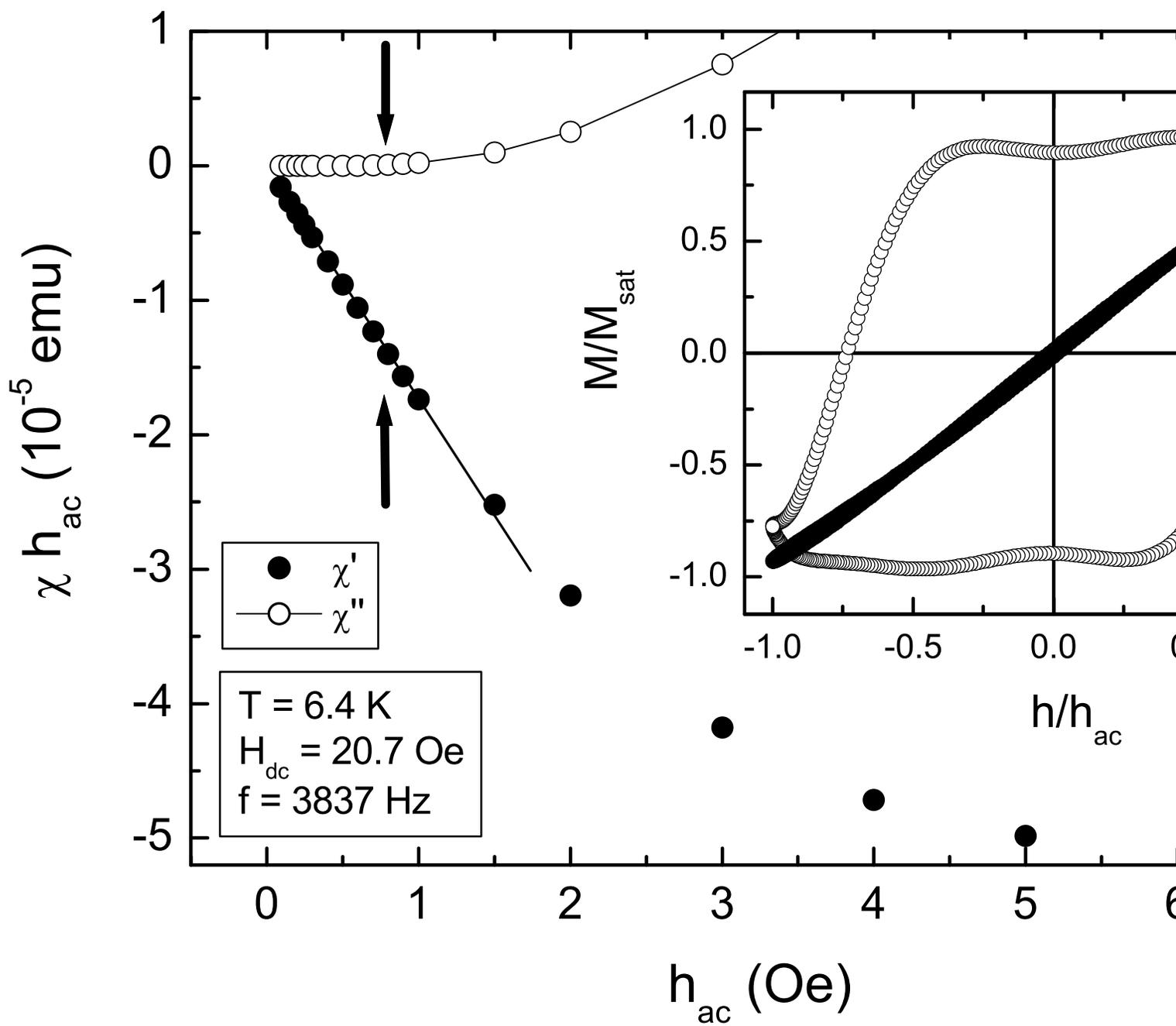
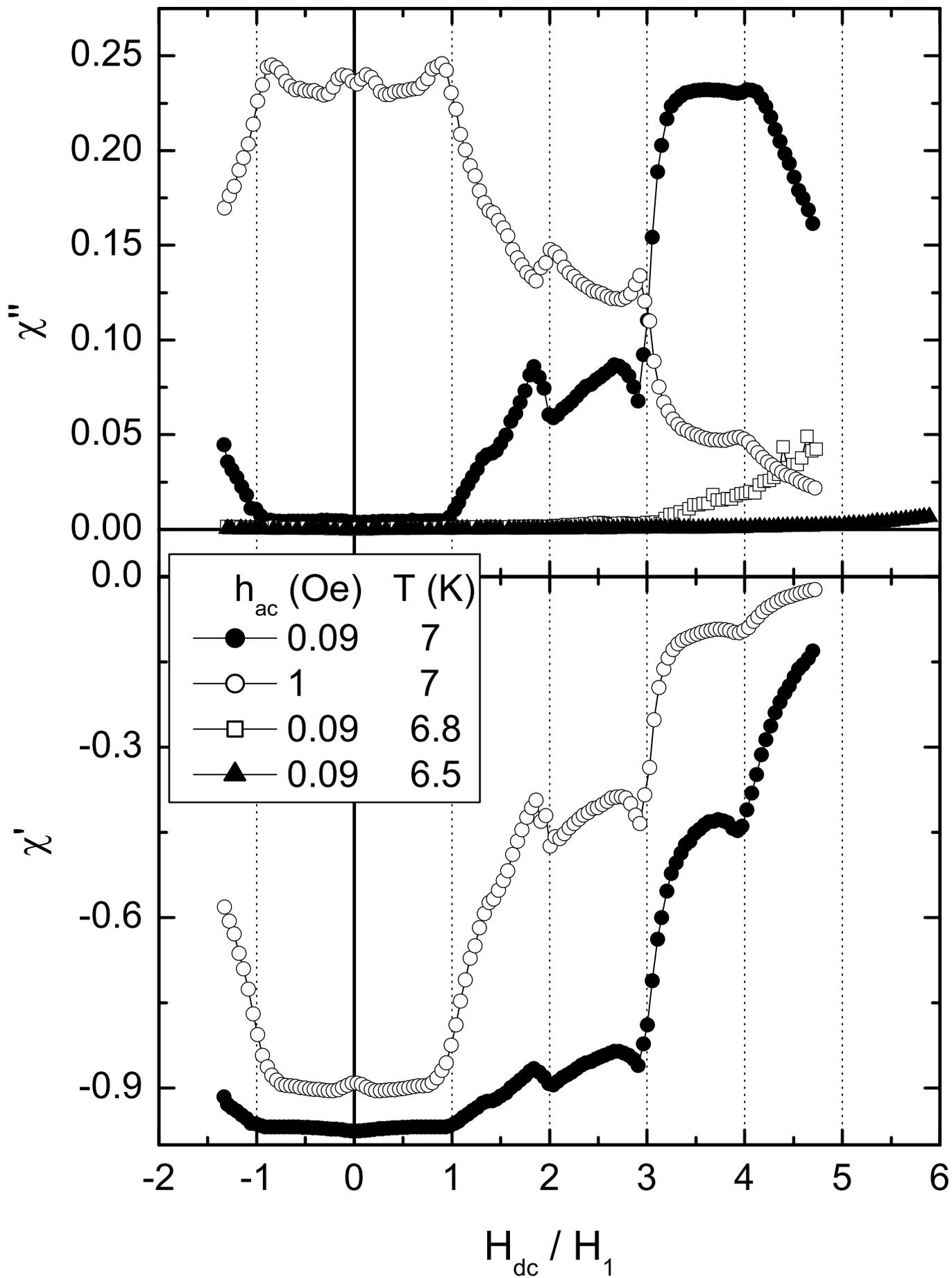
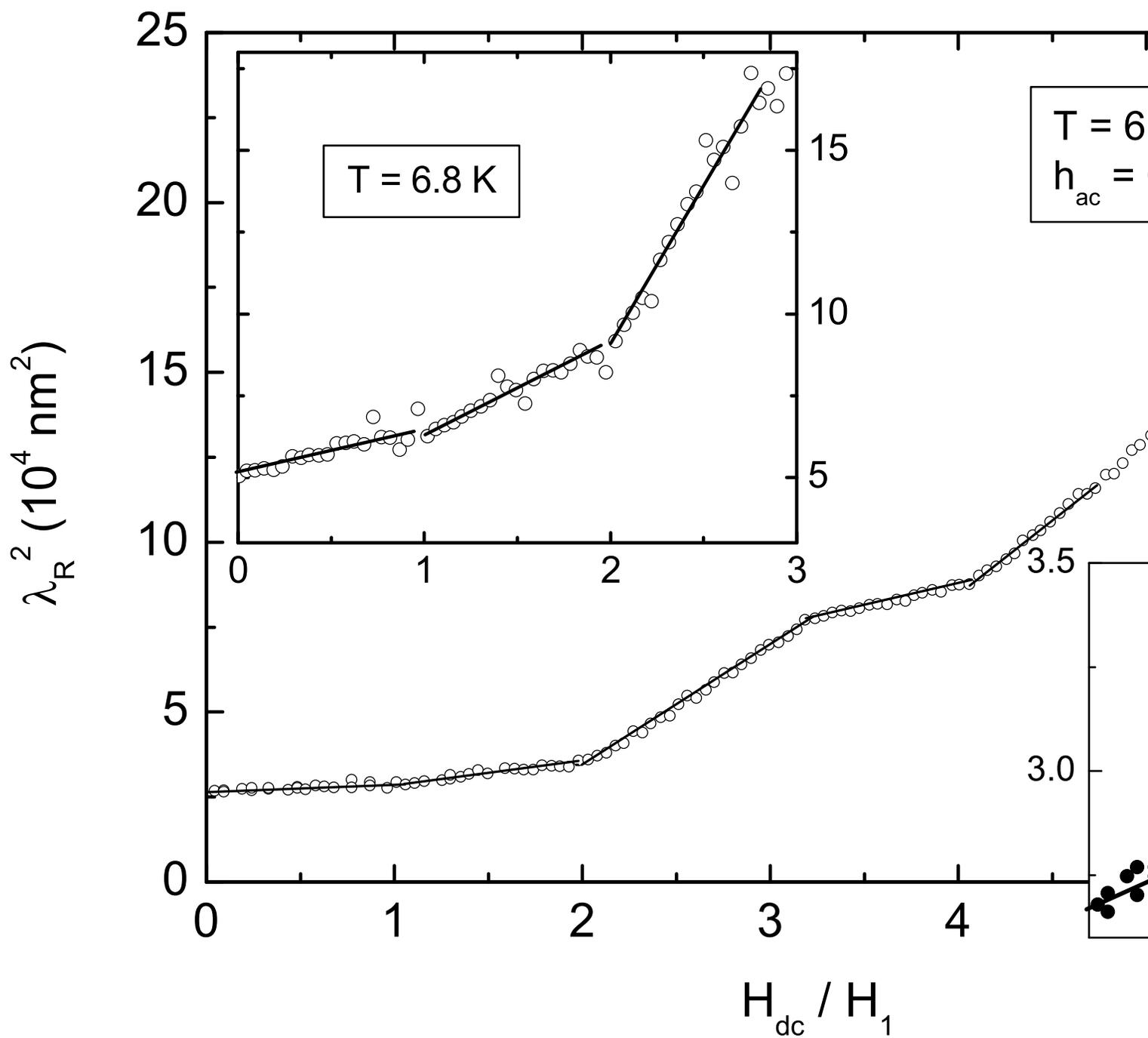
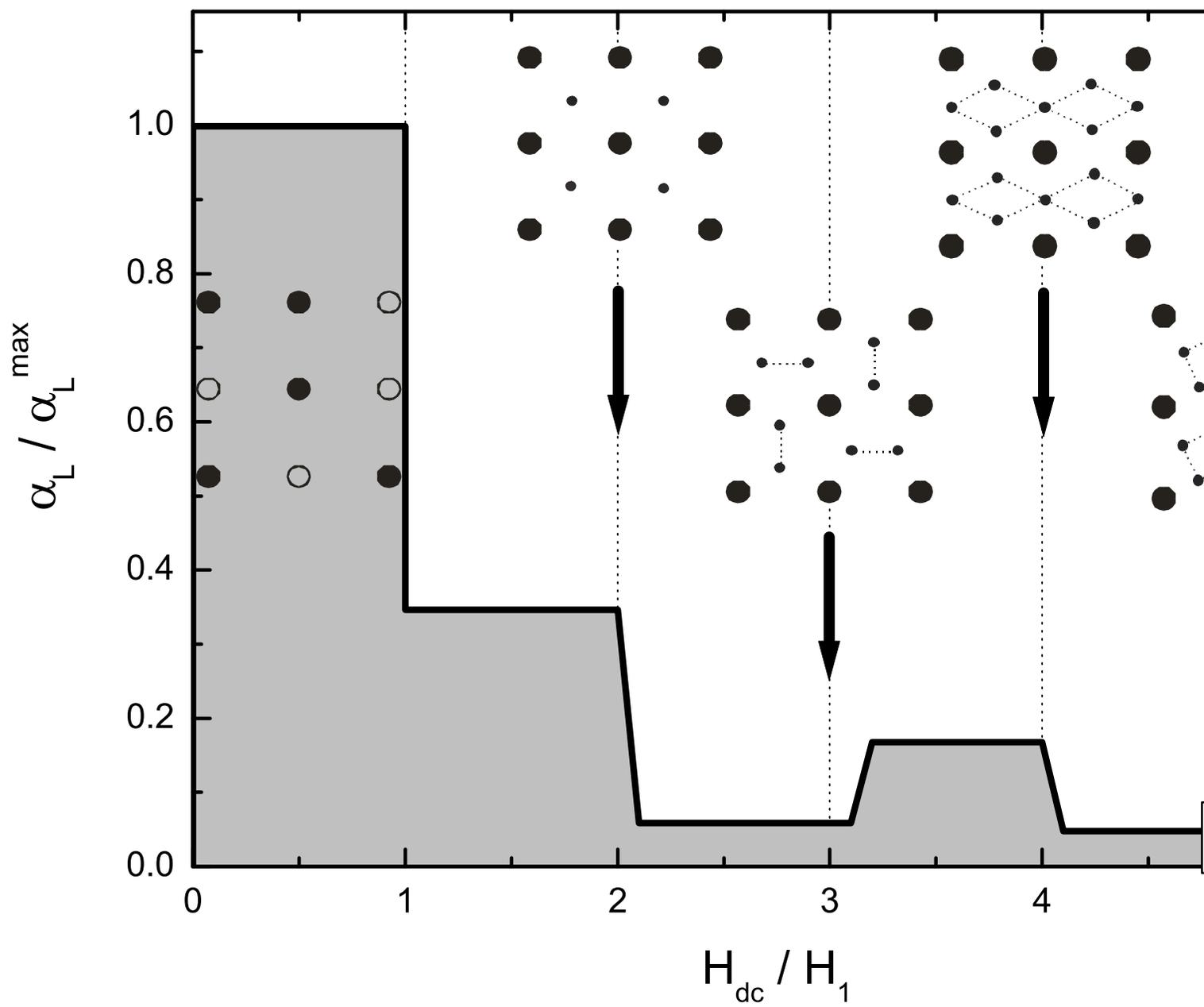
\end{document}